\documentclass[a4paper,fleqn,usenatbib,useAMS]{mnras}

\usepackage{graphicx}	
\usepackage{amsmath}	
\usepackage{amssymb}	
\usepackage{multicol}        
\usepackage{bm}		
\usepackage{pdflscape}	
\usepackage[encapsulated]{CJK}
\usepackage{ucs}
\usepackage[utf8x]{inputenc}
\usepackage{multirow}
\usepackage{booktabs,caption}
\usepackage[flushleft]{threeparttable}
\usepackage[dvipsnames]{xcolor}

\usepackage[T1]{fontenc}
\usepackage{ae,aecompl}

\usepackage{newtxtext,newtxmath}

\title[X-rays from J210204]{X-ray observations of the nova shell IPHASX J210204.7$+$471015\thanks{This work was based on observations obtained with {\it XMM-Newton}, an ESA science mission with instruments and contributions directly funded by ESA Member States and NASA.}}
\author[Toal\'{a} et al.]{{J.A.\,Toal\'{a}$^{1}$\thanks{E-mail: j.toala@irya.unam.mx}, G.\,Rubio$^{2,3}$, E.\,Santamar\'{i}a$^{2,3}$, M.A.\,Guerrero$^4$, S.\,Estrada-Dorado$^{1}$, G.\,Ramos-Larios$^{2,3}$,}
\newauthor{and L.\,Sabin$^{5}$}
\\
$^1$Instituto de Radioastronom\'{i}a y Astrof\'{i}sica (IRyA), UNAM Campus Morelia, Apartado postal 3-72, 58090 Morelia, Mexico\\
$^2$CUCEI, Universidad de Guadalajara, Blvd. Marcelino Garc\'\i a Barrag\'an 1421, 44430, Guadalajara, Jalisco, Mexico \\
$^3$Instituto de Astronom\'\i a y Meteorolog\'\i a, Dpto.\ de F\'\i sica,CUCEI, Av.\ Vallarta 2602, 44130, Guadalajara, Jalisco, Mexico\\
$^4$Instituto de Astrof\'\i sica de Andaluc\'\i a, IAA-CSIC, Glorieta de la Astronom\'\i a s/n, 18008, Granada, Spain\\
$^{5}$Instituto de Astronom\'{i}a, Universidad Nacional Autonoma de M\'{e}xico, Apartado Postal 877, 22800 Ensenada, B. C., Mexico
}

\pubyear{2020}

\begin{document}
\label{firstpage}
\pagerange{\pageref{firstpage}--\pageref{lastpage}}
\maketitle

\begin{abstract}

\noindent 
We present the analysis of {\it XMM-Newton} European Photon Imaging Camera (EPIC) observations of the nova shell IPHASX J210204.7$+$471015. 
We detect X-ray emission from the progenitor binary star with properties that resemble those of underluminous intermediate polars such as DQ\,Her: 
an X-ray-emitting plasma with temperature of $T_\mathrm{X}=(6.4\pm3.1)\times10^{6}$ K, a non-thermal X-ray component, and an estimated X-ray luminosity of $L_\mathrm{X}=10^{30}$ erg~s$^{-1}$. 
Time series analyses unveil the presence of two periods, the dominant with a period of $2.9\pm0.2$~hr, which might be attributed to the spin of the white dwarf, and a secondary of $4.5\pm0.6$~hr that is in line with the orbital period of the binary system derived from optical observations. 
We do not detect extended X-ray emission as in other nova shells probably due to its relatively old age (130--170 yr) or to its asymmetric disrupted morphology which is suggestive of explosion scenarios different to the symmetric ones assumed in available numerical simulations of nova explosions.

\end{abstract}





\section{INTRODUCTION}
\label{sec:intro}

Nova events are the result of thermonuclear explosions from binary systems that occur when a white dwarf (WD) accretes H-rich material from a companion \citep[e.g.,][]{Gallagher1978}. Due to momentum conservation, the material forms an accretion disk before being accreted by the WD. The material in the disk spirals down onto the WD while its being heated up by friction. If the accretion rate onto the WD is low, the accreted material becomes degenerate, making the pressure and temperature to increase
dramatically. An explosion will ignite the material accreted on the
surface of the WD through a thermonuclear runaway event – a nova explosion. This thermonuclear runaway releases and initial energy $E_{0}\sim10^{44}$--10$^{45}$~erg, 
ejecting $M_\mathrm{ej}\approx$10$^{-5}$--10$^{-4}$~M$_\odot$ at velocities as high as $v_\mathrm{ej}\gtrsim10^{3}$~km~s$^{-1}$ \citep[see, e.g.,][]{Bode2010,DellaValle2020} into the interstellar medium (ISM).

The violent ejection of material in the nova event sweeps, shocks and compresses the surrounding ISM. This interaction will create an adiabatic shock that heats up the gas inside the nova remnant creating a hot bubble similarly to that of SNe and wind-blown bubbles \citep{Gudel2008}. 
The post-shock temperature of this hot bubble can be estimated as $T \propto v_\mathrm{ej}^2$ \citep{Dyson1997}. The highly supersonic ejecta should produce post-shock temperatures in excess to $10^{6}$~K, that in principle could be detected with X-ray instruments. Due to the high thermal energy of the hot bubble, it will dominate the dynamical evolution of the nova remnant until its dispersal into the ISM. During its evolution, the nova shell will experience Rayleigh-Taylor instabilities and will be disrupted \citep[see, e.g.,][]{Orlando2017}; a characteristic that is evident in narrow-band imaging of novae shells \citep[see][and references therein]{Santamaria2020}. The clumps and filaments will create ablation flows \citep[e.g.,][]{Vaytet2007}, mixing the nova shell material with that of the hot bubble, reducing its temperature to $\sim10^{6}$~K enhancing the production of soft X-ray emission \citep{Toala2018}.

There are two nova remnants with unambiguously resolved extended X-ray emission and they seem to exhibit contrasting aspects of the same event. On one hand, GK\,Per can be described as a scaled-down version of a SN remnant producing its thermal and non-thermal X-ray emission as the result of the expansion of the shock wave compressing the surrounding ISM \citep{Balman1999,Balman2005,Takei2015,Yuasa2016}. Meanwhile DQ\,Her has been recently suggested to harbour the first ever detected magnetised jet disclosing more fundamental physics acting at the level of the progenitor binary star \citep{Toala2020}.

Extended X-ray emission has been suggested to be present in other nova-like systems: the classical nova RR\,Pic \citep{Balman2004}, the cataclysmic variable (CV) DK\,Lac \citep{Takei2013} and those around the recurrent novae T\,Pyx \citep{Balman2014} and RS\,Oph \citep{Luna2009}. However, their marginal detections, in some cases unresolved, make it difficult to unveil the true spatial distribution of the X-ray emission.

\begin{figure}
\begin{center}
\includegraphics[width=\linewidth]{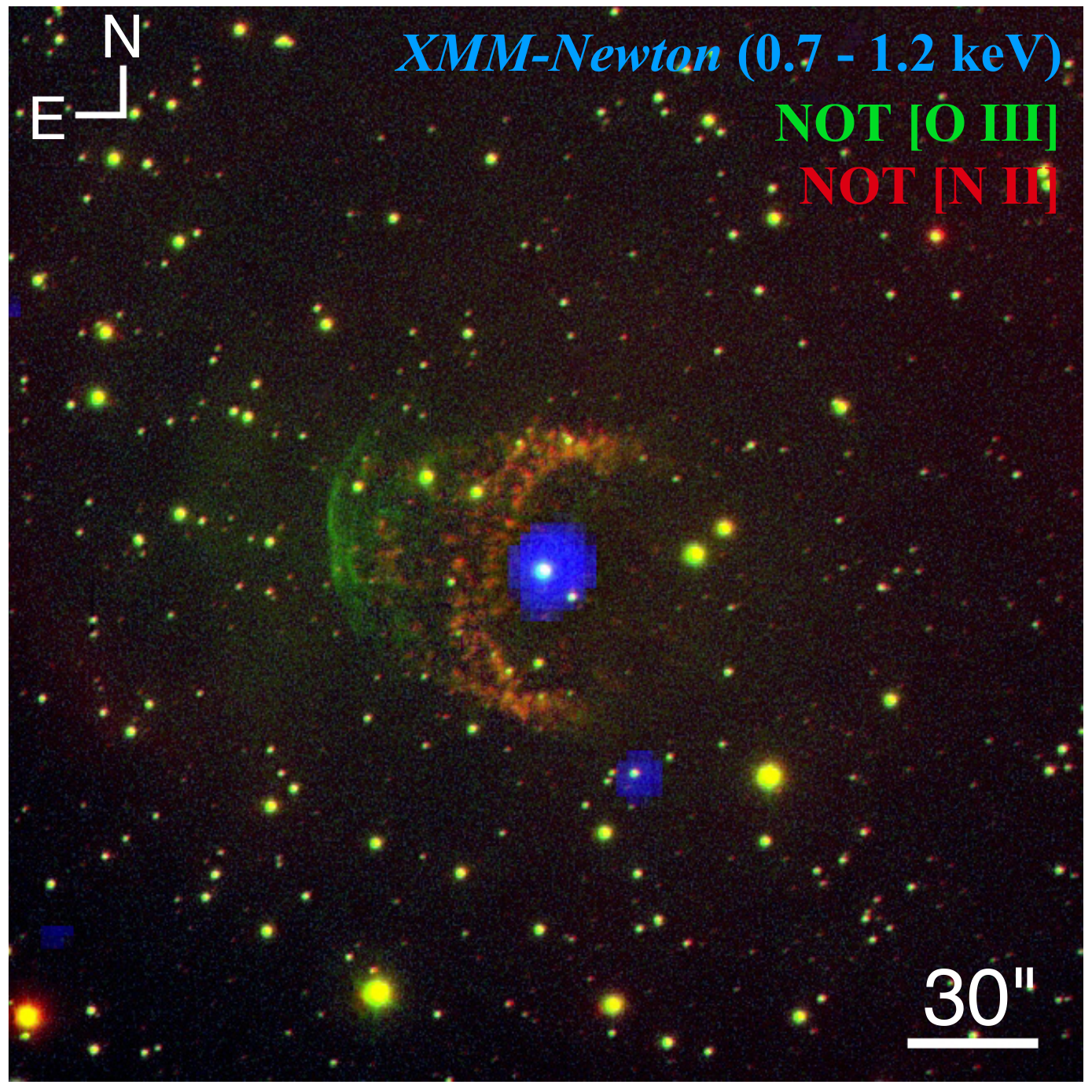}
\caption{Optical and X-ray colour-composite picture of IPHASX J\,210204.7$+$471015. 
Red and green correspond to narrow-band images obtained with the NOT \citep[see][]{Guerrero2018}, while blue corresponds to the {\it XMM-Newton} X-ray image in the 0.7--1.2~keV energy range. 
The progenitor star system is at the centre of the image. 
North is up, east to the left.
}
\label{fig:image}
\end{center}
\end{figure}

In contrast, X-ray emission from the central binary system is often reported in the literature. The evolution of the X-ray properties of the binary system in a nova eruption vary significantly during the outburst and post-outburst \citep[see, e.g.,][and references therein]{Singh2020}. In the non-magnetic cases, after accretion is reestablished, the progenitor system will behave similarly to a CV \citep{Hernanz2002,Hernanz2010,Sala2017}.
Their X-ray emission is likely to be produced on post-shock regions at the base of the accreting surface on the WD \citep{Eracleous1991,Baskill2005}. In the cases in which the WD has a strong magnetic field ($B\approx$1--10~MG), as the case of the so-called intermediate polars (IPs), it quenches the inner region of the accretion disk at the magnetospheric radius of the WD. The material falls into the WD following the magnetic field lines towards the polar regions forming a shock \citep[see][]{Patterson1994} and producing X-rays \citep[see][and references therein]{Worpel2020}. In such cases the spin of the WD is not synchronised to the orbit of the binary system, and thus, some X-ray variability is seen even in quiescent state \citep[e.g.,][]{Mukai2003}.

In this paper we present an X-ray characterisation of the nova shell IPHASX J210204.7$+$471015 (hereinafter J210204; see Fig.~\ref{fig:image}) and its progenitor system. J210204 was discovered as part of the Isaac Newton Photometric H$\alpha$ Survey \citep[IPHAS;][]{Drew2005,Barentsen2014} and proposed to be a possible planetary nebula \citep{Sabin2014}, but later characterised as a nova shell \citep{Guerrero2018}. 
The most obvious structure of J210204 illustrated in Figure~\ref{fig:image} is a clumpy incomplete ring-like structure mainly detected in [N\,{\sc ii}], with most of the knots exhibiting a cometary morphology. 
This ring-like structure lacks emission towards the west with less dense fewer knots. 
A different structure is detected in [O\,{\sc iii}], a bow-shock-like feature towards the east, which might be caused as the shock wave expands into the ISM. 
The clear asymmetric morphology of J210204, reminiscent of the classical nova AT\,Cnc \citep{Shara2012}, is not common for nova shells. Their morphology represents the best piece of evidence of asymmetric explosions shaping novae shells.

We use {\it XMM-Newton} to search and characterise the X-ray emission from its central engine and to assess the presence of a hot bubble. 
This paper is organised as follows. 
In Section~2 we present the observations and the data analysis. 
In Section~3 we present our results. 
A discussion is presented in Section~4. 
Finally, we summarise our results in Section~5.

\section{Observations and data preparation}

J210204 was observed with the three European Photon Imaging Cameras (EPIC) on board {\it XMM-Newton} during 2020 May 5 for a total observation time of 72~ks. The observations were obtained on the Extended Full Frame mode for the pn camera and in the Full Frame mode for the MOS 1 and MOS2 cameras. 
All observations were obtained with the medium optical blocking filter. 
The total exposure times on the MOS1, MOS2 and pn cameras are 67.9, 68.1 and 64.7~ks, respectively.

\begin{figure}
\begin{center}
\includegraphics[width=\linewidth]{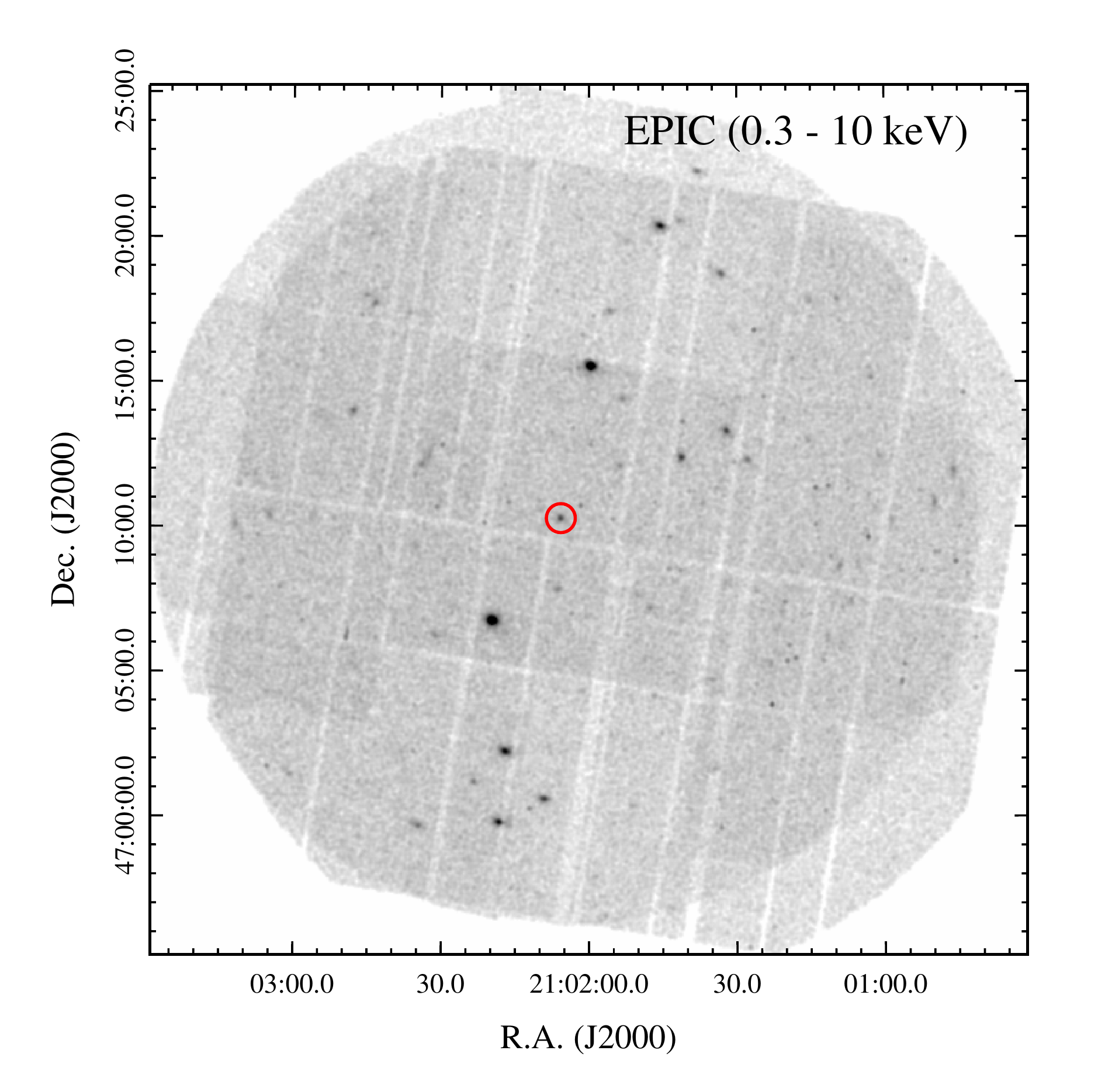}
\caption{
{\it XMM-Newton} EPIC (MOS1$+$MOS2$+$pn) image in the 0.3--10~keV energy range of IPHASX J\,210204.7$+$471015. 
The position of the progenitor star of J210204 is shown with a red circle with an angular radius of 20~arcsec.
}
\label{fig:EPIC}
\end{center}
\end{figure}

The observation data files (ODF) were processed with the Science Analysis System ({\sc sas}) version 19.0\footnote{\url{https://www.cosmos.esa.int/web/xmm-newton/what-is-sas}}. The event files were produced using the {\it epproc} and {\it emproc} {\sc sas} tasks with the calibration files obtained on 2020 November 30. In order to excise periods of high-background we created EPIC light curves binning the data over 100~s for the 10--12~keV energy range. The background was considered high for count rates  of 0.2~counts~s$^{-1}$ for the EPIC-MOS and 0.45~counts~s$^{-1}$ for EPIC-pn. 
After excising bad periods of time, the net exposures times for the MOS1, MOS2 and pn cameras resulted in 58.2, 57.9 and 40.8~ks, respectively.

The clean event files of the three EPIC cameras have been combined using the {\it merge} SAS task only for purposes of illustration. In Figure~\ref{fig:EPIC} we show this image and the position of the progenitor star of J210204. 
The central source is detected in X-rays.

\begin{figure}
\begin{center}
\includegraphics[width=\linewidth]{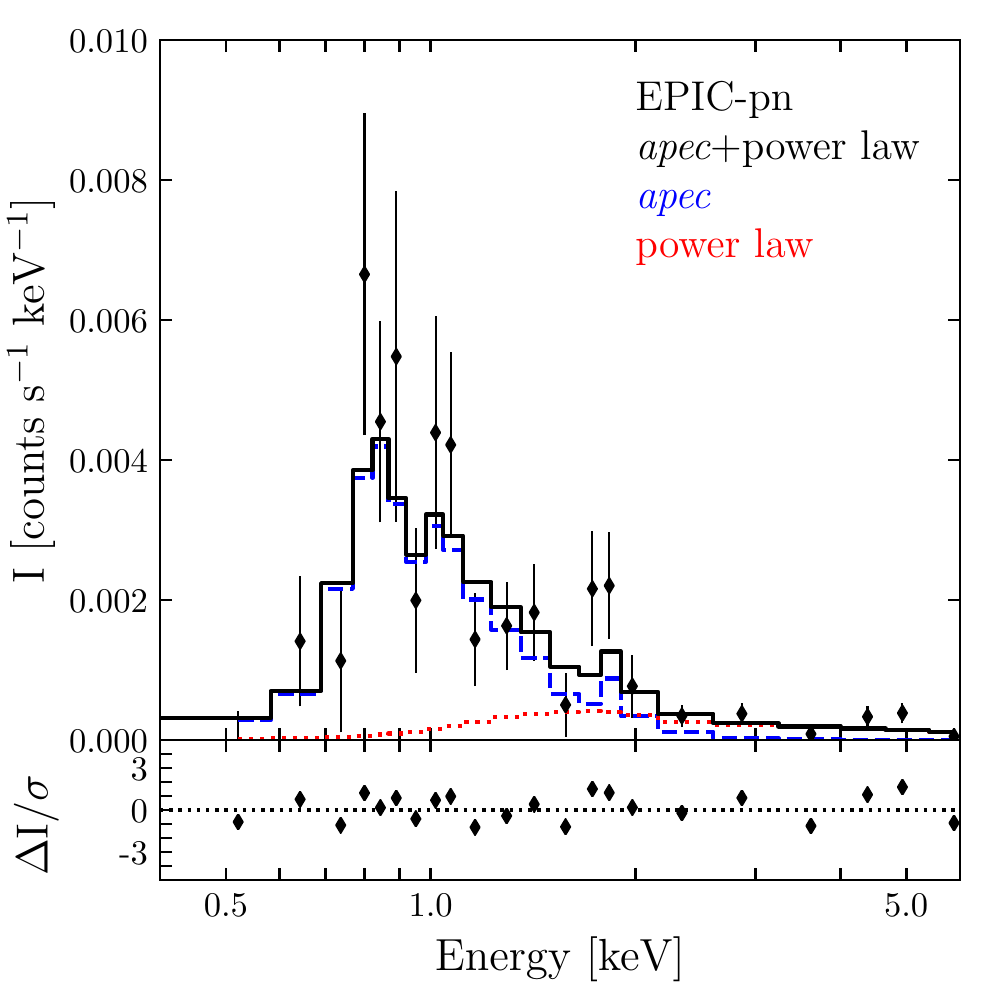}
\caption{
Background-subtracted EPIC-pn spectrum of the central source in J210204 (black diamonds). 
The solid black line represents the total model, whilst the dashed (blue) and dotted (red) lines show the contribution from the apec and power law components. 
The spectrum has been extracted requesting a minimum of 10 counts per spectral bin. 
The bottom panel presents the residuals of the best-fit model.
}
\label{fig:spec}
\end{center}
\end{figure}

In order to extract spectra of J210204, we defined a source circular aperture with a radius of 20~arcsec and a background aperture from a region with no contribution from point sources in the vicinity of J210204. 
The resultant count rates for the central star of J210204 obtained from the MOS1, MOS2 and pn cameras are 0.70$\pm$0.15, 0.60$\pm$0.17 and 3.7$\pm$0.4~counts~ks$^{-1}$ with total detected count numbers of 40$\pm$9, 32$\pm$9 and 150$\pm$16~counts for the MOS1, MOS2 and pn cameras, respectively. 
The background-subtracted EPIC-pn spectrum is presented in Figure~\ref{fig:spec}. 
We do not show the MOS spectra because their low count rates result in a limited quality. 

To further search for extended X-ray emission filling the nebular shell of J210204, we also analysed the data following the Extended Source Analysis Software ({\sc esas}) tasks \citep{Snowden2004,Kuntz2008,Snowden2008}, which are currently included as part of {\sc sas}. 
The {\sc esas} tasks apply more restrictive event rejection criteria, cleaning the data for the possible contamination from the astrophysical background, the soft proton background and solar wind charge-exchange reactions, which have important contributions at energies <1.5 keV.
Independent pn, MOS1 and MOS2 images were extracted, merged together, and corrected for exposure maps. This was done for three different energy ranges, namely, a soft (0.3--0.7~keV), a medium (0.7--1.2~keV) and a hard (1.2--5.0~keV) band.

No extended X-ray emission was detected in any of the three EPIC images produced with the {\sc esas} tasks, but only the presence of point sources in the vicinity of J210204. 
In Figure~\ref{fig:image} we compare the medium X-ray band with optical images obtained at the Nordic Optical Telescope (NOT) taken with [O\,{\sc iii}] and [N\,{\sc ii}] narrow band filters.

\section{Analysis and results}
\label{sec:ana}

The bulk of the emission in the EPIC-pn spectrum of the central source of J210204 is in the energy range between 0.4 and 2~keV (see Fig.~\ref{fig:spec}). 
The spectrum hints at the presence of three emission lines at $\sim$0.8, $\sim$1.0, and $\sim$1.8~keV. 
The first two might be attributed to Ne\,{\sc ix} and Ne\,{\sc x} emission lines, respectively, whilst the latter might be due to  Si\,{\sc xiii} emission lines. 
We note that a certain energetic continuum can be seen in the spectrum from 2.0 to 6.0~keV.

In order to estimate its physical properties, the EPIC-pn spectrum of the central star of J210204 was modelled using the X-Ray Spectral Fitting Package \citep[{\sc xspec};][]{Arnaud1996} version 12.9.1. A first model was attempted by accounting for a single optically-thin {\it apec} plasma emission model. 
This single component model reproduced the soft part of the spectrum, but it is notably deficient for the hard X-ray emission above 2.0~keV. 
Models with two-component were subsequently attempted.

Our first model included two {\it apec} plasma components with solar abundances \citep{Wilms2000} which resulted in a relatively good fit with a reduced $\chi^{2}$ of 1.22. 
The soft component has a plasma temperature of $k T_1$=0.52$^{+0.24}_{-0.27}$ keV and was fitted with a hydrogen column density $N_\mathrm{H}=(8.0^{+2.7}_{-3.6})\times10^{21}$~cm$^{-2}$. 
The later is consistent with the extinction measured in optical spectra \citep[see][]{Guerrero2018}. 
However, {\sc xspec} had troubles fitting the second component, resulting in an undefined plasma temperature $k T_2$ between 2 and 64~keV.

A better fit to the EPIC-pn spectrum was achieved by a temperature plasma emission model plus a power law component. 
This fit resulted in a similar reduced $\chi^2$ of 1.20, a column density  $N_\mathrm{H}=(8.1^{+2.5}_{-3.4})\times10^{21}$~cm$^{-2}$ and a plasma component with $kT_1=0.55^{+0.22}_{-0.27}$~keV. 
The power-law component is characterised by a photon index $\Gamma=1.0^{+1.0}_{-1.4}$. 
The {\sc xspec} normalization parameters\footnote{The normalization parameter is defined in {\sc xspec} as $A=10^{-14}\int n_\mathrm{e} n_\mathrm{H} \mathrm{d}V / 4 \pi d^2$, where $n_\mathrm{e}$ and $n_\mathrm{H}$ are the electron and hydrogen number densities and $d$ is the distance to the object. The integral is performed over the volume $V$.} of the two components are $A_\mathrm{apec} = 1.82\times10^{-5}$ cm$^{-5}$ and $A_\mathrm{pow}$=1.22$\times10^{-6}$ cm$^{-5}$.

The best-fit model resulted in an absorbed flux of $f_\mathrm{X}=(1.35\pm0.35)\times10^{-14}$ erg~s$^{-1}$~cm$^{-2}$ in the 0.4--6.0~keV energy range and corresponds to an intrinsic flux of $F_\mathrm{X}=(6.9\pm3.3)\times10^{-14}$ erg~s$^{-1}$~cm$^{-2}$. 
Taking into account a distance of 0.6~kpc estimated by \citet{Santamaria2019}, we calculated an X-ray luminosity of
$L_\mathrm{X}=(3.0\pm1.4)\times10^{30}$~erg~s$^{-1}$.

Models with varying abundances were also attempted, in particular, models with Ne and Si abundances as free parameters. 
The Ne abundances resulted in 1.2 times its solar value whilst the Si abundance converged to values $\sim$3--4 times the solar value. 
However, these models did not improved considerably the previous fit with a reduced $\chi^{2}$=1.16. 
Nevertheless, the estimated errors for these two elemental abundances were larger than the actual values. 
For this, we will keep the model described in the paragraphs above as the best-fit to the EPIC-pn spectrum.

\begin{figure}
\begin{center}
\includegraphics[width=\linewidth]{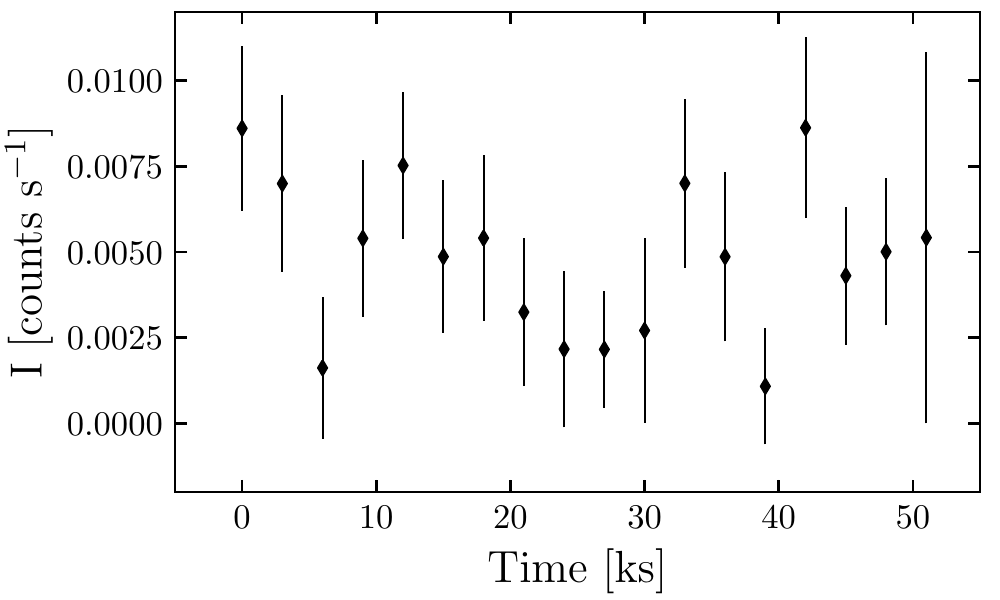}
\caption{Background-subtracted EPIC-pn light curve in the 0.3--6.0~keV energy range of the central binary system in J210204. Each bin represents 3~ks.}
\label{fig:lc}
\end{center}
\end{figure}

To assess the possible X-ray variability of the binary system of J210204, we extracted background-subtracted light curves in the 0.3--6.0, 0.4--2.0 and 1.5--6.0~keV energy ranges. 
The three light curves are very similiar and, thus, we only show that of the EPIC-pn corresponding to the 0.3--6.0~keV energy range in Figure~\ref{fig:lc}. 
We used the PyAstronomy Lomb-Scargle routine to produce a periodogram and calculate the most possible period of variation. This procedure suggests a period of 10.7$\pm$0.6~ks (=2.9$\pm$0.2~h) for the 0.3--6.0 and 0.4--2.0~keV light curves and a period of 11.1$\pm$0.7~ks (=3.1$\pm$0.2~hr) for the light curve in the 1.5--6.0~keV range, i.e., the periods in the different energy bands are consistent within their uncertainties.
We note that each periodogram exhibited the presence of other two secondary components. We illustrate this in Figure~\ref{fig:periodo} were we show the normalized power spectrum obtained for the light curve presented in Figure~\ref{fig:lc}. 
This periodogram has three main peaks at frequencies  
$\omega_1=(6.2\pm0.9)\times10^{-5}$ Hz, 
$\omega_2=(9.4\pm0.4)\times10^{-5}$ Hz and 
$\omega_3=(1.2\pm0.7)\times10^{-4}$ Hz. 
These correspond to periods of 
$P_1$=4.5$\pm$0.6 hr, 
$P_2$=2.9$\pm$0.2 hr and 
$P_3$=2.3$\pm$0.7 hr, respectively. 
$P_2$ is the one described above while $P_1$ is consistent with that estimated from optical observations of the progenitor system of J210204 of 4.26$\pm$0.01~hr \citep[see][]{Guerrero2018}.

As shown in the previous section, no extended X-ray emission is detected with the current EPIC observations. 
In order to estimate upper limits to the X-ray flux and luminosity of the possible hot bubble in J210204, we extracted a background-subtracted spectrum of a region encompassing that of the nebular emission, excising the contribution from point sources. 
We estimated a 3-$\sigma$ upper limit to the EPIC-pn count rate in the 0.3--6.0~keV of 0.54 counts~ks$^{-1}$. 
Using the count rate simulator PIMMS\footnote{\url{https://cxc.harvard.edu/toolkit/pimms.jsp}} version 4.10, assuming a plasma temperature for the hot bubble of $T=2\times10^{6}$~K and adopting the $N_\mathrm{H}=8\times10^{21}$~cm$^{-2}$ as obtained for the spectral fit of the progenitor system of J210204, we estimated an absorbed flux of $f_\mathrm{X,DIFF}\lesssim2.1\times10^{-16}$~erg~s$^{-1}$~cm$^{-2}$. This corresponds to an intrinsic flux of $F_\mathrm{X,DIFF}\lesssim2.5\times10^{-14}$ erg~s$^{-1}$cm$^{-2}$ and an upper limit to the X-ray luminosity of $L_\mathrm{X,DIFF}\lesssim1.1\times10^{30}$~erg~s$^{-1}$.

\section{Discussion}

Our best-fit model to the EPIC-pn spectrum of the progenitor system of J210204 suggests the presence of a thermal component with a temperature of $T_\mathrm{X}\approx(6.4\pm3.1)\times10^{6}$~K. 
Interestingly, the second component is best described by a power law with a spectral index of $\Gamma=1.0$, suggesting the presence of non-thermal emission an, thus, a possible IP origin for this stellar system. We note that the number of IP systems in the Galaxy are expected to be large, contributing importantly to the Galactic X-ray properties \citep{Revnivtsev2009,Warwick2011,Warwick2014}, but not many have been confirmed in the literature\footnote{The Catalogue of IPs and IP Candidates report 24 confirmed IPs by 2014 October 29. See \url{https://asd.gsfc.nasa.gov/Koji.Mukai/iphome/catalog/alpha.html}}. Thus, the characterisation of such systems as possible IPs is crucial.

The plasma temperature of the progenitor star of J210204 is very similar to that estimated for other IPs with very similar spectra with dominant emission around 1.0~keV and emission lines above 2~keV \citep[see][and references therein]{Worpel2020}. We note that although the X-ray emission from IPs is expected to have X-ray luminosities in excees to $10^{32}$~erg~s$^{-1}$ \citep[][]{Patterson1994}, the progenitor system of J210204 has a luminosity of $10^{30}$~erg~s$^{-1}$, i.e., it is underluminous as the iconic DQ\,Her \citep[see the recently characterized cases presented in][and references therein]{Nucita2020a,Nucita2020}.

\begin{figure}
\begin{center}
\includegraphics[width=\linewidth]{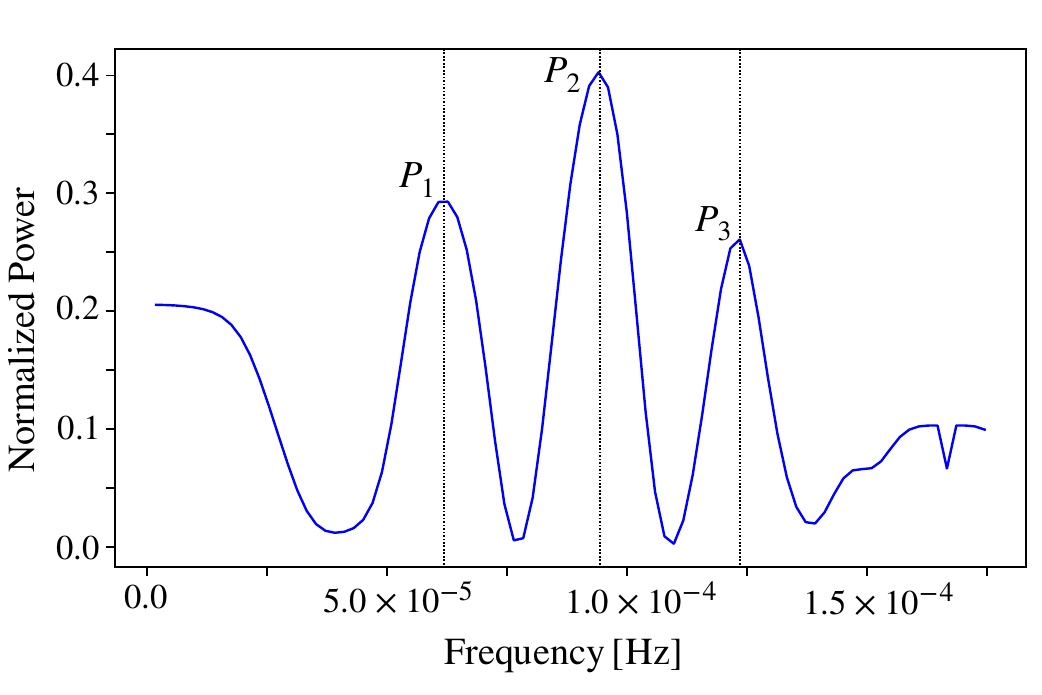}
\caption{Normalized power spectrum of the light curve presented in Figure~\ref{fig:lc}. The three peaks marked with dotted lines correspond to frequencies of $(6.2\pm0.9)\times10^{-5}$~Hz, $9.4\pm0.4\times10^{-5}$~Hz and $(1.2\pm0.7)\times10^{-4}$~Hz which corresponds to periods of $P_1$=4.5$\pm$0.6~hr, $P_2$=2.9$\pm$0.2~hr and $P_3$=2.3$\pm$0.7~hr, respectively.}
\label{fig:periodo}
\end{center}
\end{figure}

Although the resultant light curves of the EPIC observations of the progenitor system of J210204 do not have an excellent quality, we were able to perform time series analysis. The power spectrum presented in Figure~\ref{fig:periodo} exhibits three main components. 
The frequency peak $\omega_1$ implies a period of $4.5\pm0.6$~hr, which is very similar to that estimated by \citet{Guerrero2018} using optical observations. Thus, this is very likely to be the orbital period $P_\mathrm{orbit}$. If this system is indeed a IP, the dominant frequency $\omega_2$ with a period of $P_2$=2.9$\pm$0.2~hr could be attributed to the orbital spin of the WD component in the system ($P_\mathrm{spin}$). We tried to assess the origin of the third component $\omega_3$ and it seems to follow the relation
\begin{equation}
 \omega_3 =   \frac{2}{P_\mathrm{spin}} - \frac{1}{P_\mathrm{orbit}}
\end{equation}
\noindent which has been discussed thoroughly in the theoretical calculations of X-ray power spectra presented by \citet{Wynn1992}. 
These authors suggested that such frequency could be used an indicator of diskless accretion mechanisms, but this seems to depend on different parameters used for the emission model. 
We reckon, however, that our light curves and periodograms do not have sufficient quality to further discuss the origin of the accretion onto the progenitor system of J210204. 
Indeed, broad peaks can appear in the power spectrum as the result of the combination of several components in low quality X-ray light curves \citep[e.g., see figure~8 in][]{Wang2020}.


\subsection{On the lack of extended X-ray emission}

J210204 has a complex nova shell structure, unlike the ellipsoidal shell commonly observed around classical novae \citep[see, e.g.,][]{Santamaria2020}, but it shares stark similarities to the nova shell around AT\,Cnc \citep{Shara2012}. The morphological similarities between these two novae suggest that their asymmetries are not produced by interactions with the ISM, but due to the conditions of the explosion itself.

Departures from symmetry might be attributed to the interaction of the ejecta with the companion star and material in the orbital plane of the binary system, but simulations presented in \citet{Orlando2017} show that material is mostly collimated in polar directions of the orbital plane.  
The same group predicted the evolution of the X-ray luminosity for similar numerical simulations \citep{Orlando2009}. 
They predicted that the total X-ray flux in their simulations achieves a maximum of $\sim10^{-9}$~erg~s$^{-1}$~cm$^{-2}$ after the fourth day of the explosion and then decays as $\sim t^{-1.2}$ mainly due to radiative cooling. 
Assuming that such evolution of the X-ray flux is valid for J210204, their model predicts an X-ray flux of $(1.0-2.5)\times10^{-14}$ erg~s$^{-1}$~cm$^{-2}$ at the age of 130--170~yr suggested by \citet{Santamaria2020} for J210204, very similar to the upper limit estimated here using {\it XMM-Newton} EPIC-pn observations. 
Nevertheless, we would like to remark that different parameters included in the simulations (e.g., $E_{0}$ and $M_\mathrm{ej}$) could produce a range of X-ray properties \citep[e.g.,][]{Orlando2012}.

However complex the role of the different parameters explored in the simulations discussed above, all of them produce ellipsoidal morphologies \citep[see also][]{Walder2008}, which is not the case of J210204 and AT\,Cnc. 
The fact that the ejecta of these two novae appear to expand towards one side is suggestive of an explosion that originated from a specific location of the WD surface, that is, the ejecta is not only axi-symmetric, but also asymmetric. 
Future simulations adopting such scenario might be explored in the future to explain the morphologies of J210204 and AT\,Cnc.

\section{Summary}

We presented the analysis of {\it XMM-Newton} observations of the nova shell
IPHASX J\,210204.7$+$471015 (J210204). Our EPIC observations detected X-ray
emission from the central binary star with properties that are very similar to the so-called IPs: a dominant plasma temperature with temperature of $T_\mathrm{X}=(6.4\pm3.1)\times10^{6}$~K with a power-law component, signature of non-thermal X-ray emission. Its luminosity, $L_\mathrm{X}=10^{30}$~erg~s$^{-1}$, is below the expected value for the X-ray emission from IPs of 10$^{32}$--10$^{34}$~erg~s$^{-1}$, but consistent with that of DQ\,Her and other underluminous IPs.

We used time series analysis of the EPIC observations and obtained three dominant frequencies to the power spectrum. The peak with period 4.5$\pm$0.6~hr seems to correspond to the orbital period of the system as observed at optical wavelengths \citep[e.g.,][]{Guerrero2018}. 
The dominant component to the periodogram corresponds to a period of 2.9$\pm$0.2~hr very likely related to the spin of the WD. Future deeper X-ray observations are necessary to assess the presence of the dominat components in the power spectrum.

We used the ESAS tasks to search for extended X-ray emission within J210204 as signature of the presence of an adiabatically-shocked hot bubble without success. We estimated an upper limit of the flux and luminosity of the X-ray emission of $F_\mathrm{X,DIFF}\lesssim2.5\times10^{-14}$~erg~s$^{-1}$ and $L_\mathrm{X,DIFF}\lesssim1.1\times10^{30}$~erg~s$^{-1}$, respectively. This flux is very similar to that estimated from numerical simulations presented in the literature in which the explosion generates an ellipsoidal nova shell. 
However, due to the remarkable morphology of J210204, we suggest that the explosion might have not occurred isotropically at the boundary of the WD, but it would be rather asymmetric arising from a specific region of the WD. Such claim will have to be put to test with future numerical simulations that will also serve to explain the similar morphology of AT\,Cnc.

\section*{Acknowledgements}

The authors thank the referee for a prompt report.
JAT acknowledges support from the Fundaci\'{o}n Marcos Moshinsky (M\'{e}xico).  JAT is
funded by PAPIIT project IA100720 of the Direccion General de Asuntos
del Personal Academico (DGAPA) of the Universidad Nacional
Aut\'{o}noma de M\'{e}xico (UNAM). ES and GR are thankful to Consejo Nacional de Ciencia y Tecnolog\'{i}a
(CONACyT - M\'{e}xico) for their student grants. 
MAG acknowledges support of the Spanish Ministerio de Ciencia,
Innovaci\'{o}n y Universidades (MCIU) grant PGC2018-102184-B-I00 and the 
State Agency for Research of the Spanish
MCIU through the “Center of Excellence Severo Ochoa” award
to the Instituto de Astrof\'{i}sica de Andaluc\'{i}a (SEV-2017-0709).
GR-L acknowledges support from CONACyT grant 
263373. LS acknowledges support from PAPIIT-UNAM
grant IN101819. This work has made extensive use of
NASA’s Astrophysics Data System (ADS).

\section*{Data availability}

The data underlying this work are available in the article.
The reduced observations files will be shared on reasonable request to the corresponding author.


\end{document}